\documentstyle[psfig,twocolumn,prl,aps,amssymb]{revtex}
\begin{document}
\wideabs{
                     % \draft command makes pacs numbers print
\draft

\title{Two-Roton Bound State in the Fractional Quantum Hall Effect}
\author{K. Park and J. K. Jain}
\address{Department of Physics, 104 Davey Laboratory,
The Pennsylvania State University,
Pennsylvania 16802}
\date{\today}

\maketitle

\begin{abstract}

The true nature of the lowest-energy, long-wavelength neutral excitation of the 
fractional quantum Hall effect has been a long outstanding problem.  In this Letter, we 
establish that it is a two-roton bound state.  

\end{abstract}

}

%\narrowline

The neutral excitations of the fractional quantum Hall effect \cite{Tsui}
(FQHE) have attracted considerable interest in the last fifteen years since the 
initial work of Girvin, MacDonald, and Platzman (GMP) 
\cite {GMP} in which they used a single mode 
approximation (SMA) to study the excitations at Landau level filling 
$\nu=1/3$.  The long-wavelength  
limit, the topic of the present Letter,
is of particular interest because of its relevance to  
Raman scattering experiments \cite{Pinczuk1,Pinczuk2,Pinczuk3,Davies}.
In their original work, GMP raised the 
question of whether the lowest energy excitation at long wavelengths was described 
correctly by the SMA or was instead a two-roton excitation.  
Despite some further work \cite{He}, this question has
remained unresolved until now, both because of the lack of a quantitative theory of the
two-roton excitation and because the system sizes on which
exact-diagonalization studies can be performed are too small to shed meaningful 
light on long wavelength excitations.  
We wish to show that recent developments in the composite fermion theory 
have made it possible to provide a definitive answer to this question.

\begin{figure}
\centerline{\psfig{figure=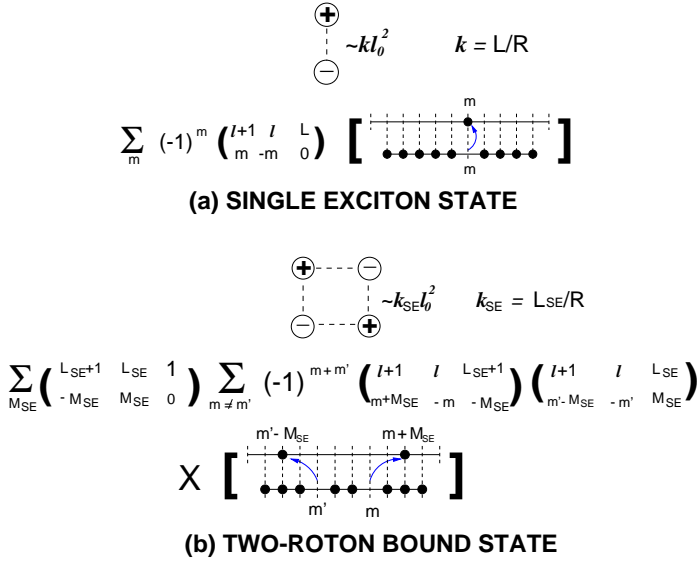,width=4.0in,angle=-90}}
\caption{Schematic diagrams for (a) the single exciton state 
and (b) the two-exciton state, with well defined $L$ and $M$ quantum numbers.
(We have chosen $M=0$ with no loss of generality, since the energy is independent 
%$\Phi_{q;L,M}^{TE}$.
of $M$.) The figure in square brackets shows schematically  
the Hartree-Fock Slater determinant obtained by promoting one or two 
electrons from the topmost occupied  Landau level to lowest 
unoccupied Landau level in single particle states indicated.  At filling factor $n$, 
the topmost occupied (lowest unoccupied) LL corresponds to the angular momentum
$l=q+n-1$ ($l=q+n$) shell in the spherical geometry; other Landau level shells are not 
shown for simplicity. The Wigner 3-j symbols are used in order to make a definite
angular-momentum eigenstate.  The relative signs of the
various terms in the sum follow from the antisymmetry requirement.
\label{fig1}}
\end{figure}

A composite fermion (CF) is the bound state of an electron and an even number of
magnetic flux quanta (a flux quantum is defined as $\phi_0=hc/e$), formed when 
electrons confined to two dimensions are exposed to a strong magnetic field
\cite{Jain,Review1,Review2}.  According to this theory, 
the interacting electrons at the Landau level (LL) filling factor 
$\nu=n/(2pn\pm 1)$, $n$ and $p$ being integers, transform into weakly
interacting composite fermions at an effective filling $\nu^*=n$; 
the ground state corresponds to $n$ filled CF Landau levels (CF-LLs) 
and it is natural to expect 
that the neutral excitations correspond to a particle-hole pair of 
composite fermions,  called the CF exciton (Fig.~\ref{fig1}a)
\cite{Kamilla,Scarola}.  
At the minimum in its dispersion, the CF exciton is called the roton, borrowing 
the terminology from the $^4$He literature \cite{GMP}.

We will use the spherical geometry \cite {Monopole} below, which considers $N$
electrons on the surface of a sphere in the presence of a radial magnetic field emanating
from a magnetic monopole of strength $Q$, which corresponds to a total flux of $2Q\phi_0$
through the surface of the sphere.
The wave function for the CF ground state or the CF exciton at $\nu=n/(2pn+1)$, 
denoted by $\Psi$, is constructed by analogy to the wave function of
the corresponding electron states at $\nu=n$, denoted by $\Phi$:
\begin{equation}
\Psi={\cal P}_{LLL} \Phi_{1}^{2p}\Phi
\end{equation}
Here $\Phi_1$ is the wavefunction of the fully occupied
lowest Landau level with monopole strength equal to $(N-1)/2$, 
given by $\prod_{j<k}(u_j v_k - u_k v_j)$,
with $u_j \equiv \cos(\theta_j /2)
\exp(-i\phi_j /2)$ and  $v_j \equiv \sin(\theta_j /2) \exp(i\phi_j /2)$.
${\cal P}_{LLL}$ denotes projection of the wave function into the lowest 
Landau level (LLL).  
The monopole strengths for $\Phi$ and $\Psi$, $q$ and $Q$, respectively, are related 
by $Q=q+p(N-1)$. For the ground state and the single exciton, the wave functions 
$\Phi$ are completely determined by
symmetry (i.e., by fixing the total orbital angular momentum $L$, which is preserved in 
going from $\Phi$ to $\Psi$ according to the above rule), giving parameter-free
wave functions $\Psi$ for the ground and excited states of interacting electrons.
These have been found to be extremely accurate in tests against exact
diagonalization results available for small systems \cite {Jain,JK}, 
establishing the essential validity of the CF exciton description of the
neutral mode of the FQHE.
However, these small system studies do not test the long wavelength limit, and there
are indications that the single exciton might not be the lowest energy excitation in the
long-wavelength limit.  First,  the agreement between the exact eigenenergy and 
the CF-exciton energy worsens somewhat at small
$k$, indicating the possibility of new physics here.  
(The wave vector $k$ is related to
the orbital angular momentum $L$ in the spherical geometry as $k=L/R$.)
Secondly, ${\cal P}_{LLL}$ projects away  
an increasingly bigger fraction of the single-exciton wave function 
as $k\rightarrow 0$,\cite{Kamilla} eventually annihilating it  completely. 
(This happens precisely at $L=1$ in the spherical geometry.\cite{Annihilation})  
Furthermore, in the 
thermodynamic limit, the energy of the single CF exciton in the $k\rightarrow 0$ limit
is slightly more than twice the energy of the CF roton, which raises the question 
of whether the two-roton excitation wins in the long wavelength limit. 
These observations have motivated us to
look into the $k\rightarrow 0$ limit in more detail.

We start by constructing a wave function $\Psi$ for an excitation consisting of two
CF excitons, for which we appeal to the analogy to the two-exciton state 
at $\nu=n$, which contains two particle-hole pairs, as shown in Fig.~(\ref{fig1}).
There are certain technical difficulties that one encounters in going from $\Phi$ to $\Psi$.
First, the final wave function is not automatically orthogonal to the {\em single} 
exciton wave function or the ground state wave function at the same $L$.  In order to 
avoid the need for complicated Gram-Schmidt orthogonalization, 
we exploit the fortunate coincidence that there is no
single exciton state at $L=1$ in the spherical geometry \cite{Annihilation}.
Therefore, we construct the two-roton wave
function at $L=1$ and compare it with the single exciton at $L=2$, 
both of which
correspond to $k\rightarrow 0$ in the thermodynamic limit. 
The second difficulty is that the wave function of the two-exciton is not unique at $L=1$.
A large number of combinations of two single excitons give a state at $L=1$:
one could combine either two single 
excitons with the same angular momenta or angular
momenta differing by unity.  We have discovered in our calculations that, 
in the former case, the LLL projection operator annihilates $\Psi$. Therefore, 
we construct $\Phi^{TE}_{q;L,M}$, the two-exciton state at $q$ with quantum numbers
$L=1$ and $M=0$, 
from two single excitons at $L_{SE}$ and $L_{SE}+1$ as shown in Fig.~(\ref{fig1}).
The two-CF-exciton state at $Q=q+p(N-1)$, $\Psi^{TE}_{Q;L,M}$, is then constructed as:
\begin{equation}
\Psi^{TE}_{Q;L,M}= {\cal P}_{LLL} \Phi_{1}^{2p}
\Phi^{TE}_{q;L,M}
\end{equation}

One of the most challeging aspects of the desired computation
is that the two-exciton CF state requires
a very large number of Slater determinants; the number 
increases as $N^3$, $N$ being the number of electrons.
For efficient Monte Carlo simulations we have devised a determinant-updating 
technique, generalizing the technique in Ref.\cite{Ceperley}. 
The updating technique begins with the observation that
the constituent Slater determinants of the two-exciton 
state differ from
the ground state only in \emph{two rows}.  To elaborate on 
our updating method, first let $[Y]^{gs}_{\alpha,j}$
denote an element of the matrix for the Slater determinant 
describing the ground state at q, i.e. $\Phi_{q}^{gs}$. Here
$\alpha$ collectively indicates orbital quantum numbers. 
Also, let $\Phi_{m,m+M_{SE}}^{m',m'-M_{SE}}(q,n)$
be the Hartree-Fock wavefunction obtained by promoting
two composite fermions in the $m$ and $m'$ state of the 
topmost occupied CF-LL ($(n-1)$th CF-LL) to the $m+M_{SE}$ 
and $m'-M_{SE}$
state of the lowest unoccupied CF-LL ($n$-th CF-LL)
which is depicted as the figure enclosed by 
square brackets in (b) of Fig.~\ref{fig1}. Then,
$\Phi_{m,m+M_{SE}}^{m',m'-M_{SE}}(q,n)$ is given by:

\begin{eqnarray}
\Phi_{q}^{gs}
&\Bigg[& \Big( \sum_{j} Y_{q,n,m+M_{SE}}(\Omega_j) 
{\overline{[Y]}}^{gs}_{(q,n-1,m),j} \Big)
\nonumber
\\
&\times&\Big( \sum_{j} Y_{q,n,m'-M_{SE}}(\Omega_j) 
{\overline{[Y]}}^{gs}_{(q,n-1,m'),j} \Big)
\nonumber
\\
&-&\Big( \sum_{j} Y_{q,n,m+M_{SE}}(\Omega_j) 
{\overline{[Y]}}^{gs}_{(q,n-1,m'),j} \Big)
\nonumber
\\
&\times&\Big( \sum_{j} Y_{q,n,m'-M_{SE}}(\Omega_j) 
{\overline{[Y]}}^{gs}_{(q,n-1,m),j} \Big)
\Bigg]
\label{eq:updating}
\end{eqnarray}
where ${\overline{[Y]}}^{gs}$ is the transpose of the inverse
matrix of ${[Y]}^{gs}$. Eq.~(\ref{eq:updating}) reduces
the number of operations by approximately a factor of $N$,
compared to that of operations when the Slater determinants
are computed separately.  Then, using parallel computing techniques,
such as message-passing interface (MPI), which reduces the (wall) computing time
by another factor of 30 with as many as 64 processors, we are able to study systems 
with up to $N=30$ particles, which will be crucial in what follows.  

\begin{figure}
\centerline{\psfig{figure=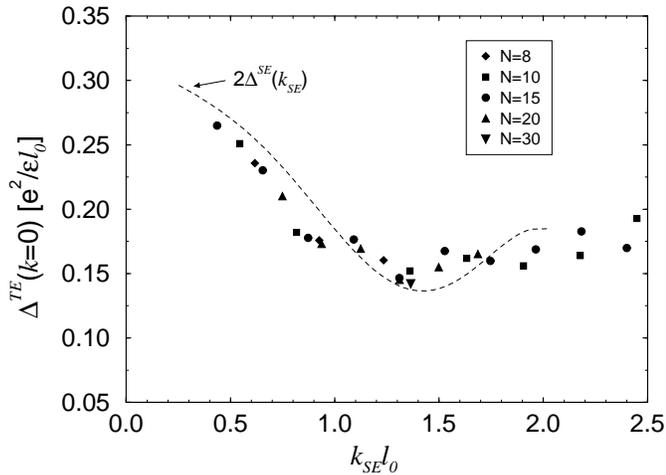,width=4.0in,angle=-90}}
\caption{The energies of various two-exciton states as a function
of the wave vector of two constituent single excitons, which is
denoted by $k_{SE}$. Here the interaction between electrons
is taken as the Coulomb interaction.
Comparison with twice the single exciton
dispersion curve (dashed line)
shows that the lowest-energy 
two-exciton state is obtained by combining two \emph{rotons}.  
\label{fig2}}
\end{figure}

The energies of the various two-exciton states 
(labeled by $L_{SE}$, or $k_{SE}=L_{SE}/R$)   
are now evaluated for the pure Coulomb interaction 
by the Monte Carlo method, treating the LLL projection operator in 
the standard method \cite{JK}.  The plot in Fig.~\ref{fig2} 
explicitly confirms that  
the lowest energy for each $N$ is obtained by combining two {\em rotons}.
In principle, the energy of the two-roton state 
could be further lowered by allowing it to mix
with other two-exciton states, but this possibility will not be considered here.  
We have compared the energy of the two-roton state against exact 
diagonalization studies to ascertain the validity of the wave function.  For 
$N=8$ and 10, energies per particle for the $L=1$ two-roton states in the 
composite fermion theory are $-0.417440 (46) $ and $-0.416551(59)$ in 
units of $e^2/\epsilon l_0$,
which are approximately 0.2\% larger than the exact energies
(-0.418324 for $N=8$ and -0.417516 for $N=10$).  
The CF energy gap at $L=1$ is approximately 5\% higher than the exact value.

\begin{figure}
\centerline{\psfig{figure=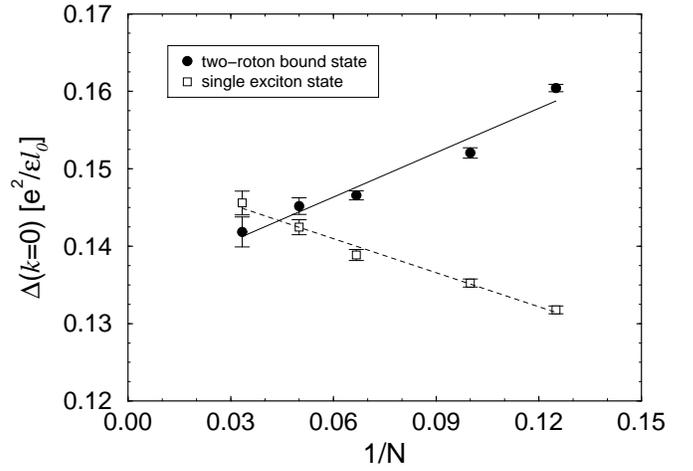,width=4.0in,angle=-90}}
\caption{The Coulomb energies as a function of $1/N$ for  
the single exciton and the two-roton bound state in the 
long wavelength limit. 
%Thermodynamic limits are computed by extrapolating
%finite-sytem energies as $1/N \rightarrow 0$, where $N$
%is the number of electrons.
\label{fig3}}
\end{figure}

Fig.~\ref{fig3} shows the evolution of the energies of the single exciton and 
the two-roton bound state as a
function of $N$.  This plot demonstrates the principal result of the work, namely that the
lowest energy excitation in the long wavelength limit is the two-roton state.
We note here that in the long wavelength limit, the single CF exciton 
is identical to the SMA mode \cite{Kamilla}, 
consistent with the fact in Fig.~\ref{fig3} that 
the energy of the single exciton approaches 0.15
$e^2/\epsilon l_0$ in the thermodynamic limit, which is also the $k=0$ energy of the 
SMA mode \cite{GMP}.  The two-roton state has 10\% lower energy.

We next come to the relevance of this work to Raman 
scattering experiment.  Our theoretical understanding of the scattering cross 
section of the FQHE gap modes observed in 
Raman scattering is rather unsatisfactory, even ignoring the complications introduced 
by the resonant nature of Raman scattering \cite {DasSarma,He}.
However, as noted in the early literature \cite{Pinczuk1}, there is 
reason to expect that the two-roton mode might couple
more strongly to light in Raman scattering than the single exciton mode: the
scattering cross section for the single exciton vanishes rapidly with the wave vector
as a result of Kohn's theorem, but there is no reason for it to 
vanish for the two-roton mode.   Also note that while the two-exciton states form a 
continuum, the two-{\em roton} states provide a peak in the density of states at the 
lower edge of the continuum.

\begin{figure}
\centerline{\psfig{figure=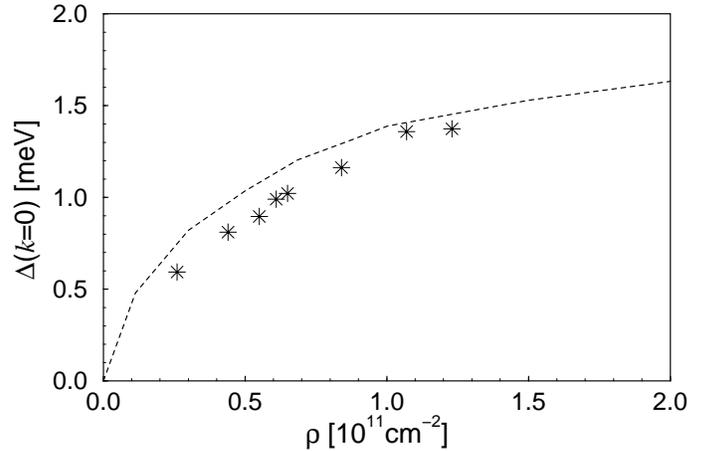,width=4.0in,angle=-90}}
\caption{Comparison between the experimetal
data (stars) from Kang {\em et al.} \protect\cite{Pinczuk4}
and the theoretical estimation of two-roton bound state energy (dashed line). 
Theoretical estimates are obtained by considering
Landau level mixing as well as finite thickness effects.  
\label{fig4}}
\end{figure}

In a detailed study, Scarola {\em et al.}\cite{Scarola} have found that, 
after incorporating the finite thickness effect, 
the energy of the single CF-roton is in excellent agreement with experiment, 
but the energy of the mode observed in the 
long wave length limit is approximately 30\% below the theoretical energy of the single 
exciton.  In light of the above discussion, it is natural to suggest that a part of  
this discrepancy might originate from a misassignment
of the nature of the long wavelength mode.  In order to make contact with experiment at 
a quantitative level, we have considered  
a square quantum well (SQW) of width 33 nm, as appropriate for the experiment of
Kang {\em et al.} \cite{Pinczuk4}  The transverse wave function has been calculated in a 
self-consistent local density approximation, as a function of the two-dimensional 
electron density, following which the integration over the transverse 
coordinate has been performed explicitly to obtain an effective two-dimensional 
interaction between electrons \cite{Stern,ParkLDA}.  This, in turn, is used to compute 
the energy of the two-roton bound state.  The thermodynamic limit 
is obtained at each density.
The excitation energies are further reduced due to Landau level mixing, which was
estimated by Scarola {\em et al.} as a function of the density
for the single roton at $\nu = 1/3$, following earlier work \cite{Bonesteel}.  
Assuming that the percent reduction of the two-roton energy is approximately 
the same (approximately 5\% for typical densities), we have obtained a realistic estimate 
for the energy of the two-roton mode, plotted (dashed line) 
in Fig.~\ref{fig4} along with the experimentally 
determined energies \cite {Pinczuk4} of the long-wavelength mode
(stars).  The agreement is on the level of 20\% or better, which we consider satisfactory
in view of various approximations and the neglect of disorder.

Our calculations also 
show a level crossing between the two-roton and the single exciton modes 
at approximately $\rho=3.0\times 10^{11}$ cm$^{-2}$ for the SQW sample of width 33nm.
If only the two-roton mode is observable in Raman scattering, this would imply that 
the Raman peak ought to disappear at sufficiently large densities.  The actual 
density at which the level crossing is predicted may be a quite sensitive 
function of the various approximations of the theory, however.

In summary, we have proved that at $\nu=1/3$ the lowest energy excitation in the 
long wavelength limit is not a single exciton but rather a two-roton mode. 
The results of Ref.~\cite{Scarola} show that much of this physics is to be expected at other
filling factors as well, since there also 
the $k\rightarrow 0$ limit of the single exciton energy is
larger than twice the single roton energy.  Calculations are in progress to investigate this
issue further.  This work 
was supported in part by the National Science Foundation under grant no. 
DMR-9986806.  We are grateful to Vito Scarola for numerous helpful discussions and 
the Numerically Intensive Computing Group led by V. 
Agarwala, J. Holmes, and J. Nucciarone, at the Penn State University CAC for 
assistance and computing time with the LION-X cluster.

\begin{references}


\bibitem{Tsui} D.C. Tsui, H.L. Stormer, and A.C. Gossard, Phys
Rev. Lett. {\bf 48}, 1559 (1982)

\bibitem{GMP} S.M. Girvin, A.H. MacDonald, and P.M. Platzman, Phys. Rev.
Lett. {\bf 54}, 581 (1985); Phys. Rev. B {\bf 33}, 2481 (1986).

\bibitem{Pinczuk1} A. Pinczuk {\em et al.}, Phys. Rev. Lett. {\bf 70}, 3983
(1993); Semicond. Sci. Technol. {\bf 9}, 1865 (1994).

\bibitem{Pinczuk2} A. Pinczuk {\em et al.}, Proceedings
of 12th Int. Conf. on High Magnetic Fields in Physics of Semiconductors, World
Scientific, 1997, p. 83.

\bibitem{Pinczuk3} Moonsoo Kang, A. Pinczuk, B.S. Dennis, M.A. Eriksson, L.N. Pfeiffer,
and K.W. West,  Phys. Rev. Lett. {\bf 84}, 546 (2000).

\bibitem{Davies} H.D.M. Davies et al., Phys. Rev. Lett. {\bf 78}, 4095 (1997).

\bibitem{He} P.M. Platzman and S. He, Phys. Rev. B {\bf 49}, 13674 (1994);
Physica Scripta {\bf T66}, 167 (1996).

\bibitem{Jain}  J.K. Jain, Phys. Rev. Lett. {\bf 63}, 199 (1989);
Phys. Rev. B {\bf 41}, 7653 (1990); J.K. Jain and R.K. Kamilla
in \cite {Review1}.

\bibitem{Review1} {\em Composite Fermions}, edited by Olle Heinonen
(World Scientific, New York, 1998).

\bibitem{Review2} {\em Perspectives in Quantum Hall Effects}, edited by S. Das
Sarma and A. Pinczuk (Wiley, New York, 1997).

\bibitem{Kamilla} R.K. Kamilla, X.G. Wu, and J.K. Jain, Phys. Rev. B {\bf 54},
4873 (1996).

\bibitem{Scarola} V.W. Scarola, K. Park, and J.K. Jain, cond-mat/9910491, 
Phys. Rev. B, in press.

\bibitem{Monopole} For the details about spherical geometry
and monopole harmonics, see 
T.T. Wu and C.N. Yang, Nucl. Phys. B {\bf 107}, 365 (1976);
F.D.M. Haldane, Phys. Rev. Lett. {\bf 51}, 605 (1983).

\bibitem{JK} J.K. Jain and R.K. Kamilla, Int. J. Mod. Phys. B {\bf 11}, 2621
(1997); Phys. Rev. B {55}, R4895 (1997).

\bibitem{Annihilation} The annihilation of the single CF exciton
at $L=1$ was discovered by G. Dev and J.K. Jain, Phys. Rev. Lett. {\bf 69}, 2843 (1992),
and has been proved analytically by K. Park and J.K. Jain, unpublished.
For the SMA wave function at $L=1$ see, S. He, S.H. Simon, and B.I. Halperin,
Phys. Rev. B {\bf 50}, 1823 (1994); 

\bibitem{Ceperley} D. Ceperley, G.V. Chester, M.H. Kalos, 
Phys. Rev. B {\bf 16}, 3081 (1977); S. Fahy, X.W. Wang, and S.G. Louie,
Phys. Rev. B {\bf 42}, 3503 (1990).
 
\bibitem{DasSarma} S. Das Sarma and Daw-Wei Wang, Phys. Rev. Lett. {\bf 83}, 816 (1999).

\bibitem{Pinczuk4}  
Moonsoo Kang {\em et al.}, to be published;
A. Pinczuk, B.S. Dennis, L.N. Pfeiffer, and K.W. West,
Physica B {\bf 249}, 40 (1998).

\bibitem{Stern} F. Stern and S. Das Sarma, Phys. Rev. B {\bf 30}, 840 (1984);
S. Das Sarma and F. Stern, {\em ibid.} {\bf 32}, 8442 (1985).

\bibitem{ParkLDA} K. Park, N. Meskini, and J.K. Jain,
J. Phys. Condens. Matter {\bf 11}, 7283 (1999), and the
references therein.

\bibitem{Bonesteel} V. Melik-Alaverdian and N.E. Bonesteel, Phys. Rev. B {\bf
52}, R17032 (1995);  R. Price and S. Das Sarma, Phys. Rev. B {\bf 52}, 17032
(1995).

\end{references}
\end{document}